\documentclass[twoside]{article}
\usepackage{fleqn,espcrc2}

\usepackage{graphicx}

\usepackage[figuresright]{rotating}

\title{Bursts of Gravitational Radiation from Superconducting Cosmic Strings and the Neutrino Mass Spectrum}

\author{Herman J.  Mosquera Cuesta\address{Abdus Salam International Centre for Theoretical Physics, Strada Costiera 11, Miramare 34014, Trieste, Italy \\ 
Centro Brasileiro de Pesquisas F\'{\i}sicas, Laborat\'orio de Cosmologia e F\.sica Experimental de Altas Energias \\ Rua Dr. Xavier Sigaud 150, Cep 22290-180, Urca, Rio de Janeiro, RJ, Brazil}\thanks{Centro Latinoamericano de F\'{\i}sica (CLAF), Avenida Wenceslau Bras 173, Botafogo, Rio de Janeiro, RJ, Brazil} and Danays Morej\'on Gonz\'alez\address{Pontif\'{\i}cia Universidade Cat\'olica do Rio de Janeiro, Rua Marqu\^es de S\~ao Vicente 225, Cep 22453-900, G\'avea, Rio de Janeiro, RJ, Brazil}} 

\begin{document}

\begin{abstract} 
{Berezinsky, Hnatyk and Vilenkin showed that superconducting
cosmic strings could be central engines for cosmological gamma-ray
bursts and for producing the neutrino component of ultra-high energy
cosmic rays. A consequence of this mechanism would be that a detectable
cusp-triggered gravitational wave burst should be released
simultaneously with the $\gamma$-ray surge. If contemporary
measurements of both $\gamma$ and $\nu$ radiation could be made for any
particular source, then the cosmological time-delay between them might
be useful for putting unprecedently tight bounds on the neutrino mass
spectrum. Such measurements could consistently verify or rule out the
model, since strictly correlated behaviour is expected for the duration
of the event and for the time variability of the spectra. 

\centerline{Published in Physics Letters B 500 (2001) 215-221}}
\end{abstract}

%\vskip 2pc
%\narrowtext 
%\twocolumn
 
%]

%\begin{keyword}: Cosmic strings -- gravitational waves -- cosmic rays -- %elementary particles: neutrino -- Detectors: LIGO, LISA, TIGAs, SK, AMANDA, BATSE-GLAST 

%\PACS 04.30.Db, 04.40.Dg,  04.50.+h 
%\end{keyword}
%\end{frontmatter}

\maketitle

\def\be{\begin{equation}} 
\def\ee{\end{equation}}

Cosmic strings (CSs) are topological defects formed during phase
transitions in the early Universe induced by spontaneous symmetry breaking
at GUT-scale energies\cite{vilenkin81}.  The energy of the unbroken vacuum
phase is released as GUT quanta of the gauge and scalar fields, forming
the CSs\cite{kibble94}.  It has been suggested that ordinary CSs could be
the cosmological sources of the biggest explosions in the
Universe\cite{MB92-93}:  the cosmological (classical) gamma-ray bursts
(GRBs)\cite{BHV2000}, with energies $E_{GRBs} \sim 10^{53-54}$ erg,
timescales $10^{-2} \leq T_{GRBs}\leq 10^3$s, as observed by
BATSE\cite{meegan95}. CSs may also be the origin of the ultra high energy
cosmic rays (UHECRs) with energies above the Greisen-Kuzmin-Zatsepin (GZK)
cut-off: $E_\times \sim 10^{19}$ eV\cite{greisen}, as well as the very
high energy neutrinos observed today\cite{kibble94,bhatta98,totani99}.  
Also, ordinary cosmic strings are potential sources of gravitational
radiation.  Emission of gravitational waves (GWs) is considered the main
channel for CS loops to decay\cite{allen00,vilenkin81}.

Turning attention to superconducting cosmic string (SCCS) loops, Babul,
Paczy\'nski and Spergel\cite{babul87}, and most recently Berezinsky,
Hnatyk and Vilenkin (BHV2000)\cite{BHV2000}, have proposed that such
objects could be the central engines of most cosmological bursts of
gamma-rays. In the first study, the currents are thought of as being
induced in the strings by primordial magnetic fields, and the source
distance scale is assumed to be $(10^2\leq z \leq 10^3)$\cite{babul87}.  
In the second one, on the other hand, the string currents are seeded by an
intergalactic magnetic field, with the GRB sources being located at
distances characteristic of superclusters of galaxies, i.e., $z \leq
5$\cite{BHV2000}.  In both of the models, a surface defect referred to as
a {\it cusp} is the trigger of the bursts, while the electric currents are
induced by oscillation of string loops in an external intergalactic
magnetic field.

A further by-product of these pictures is that a beamed surge develops
naturally at the place where a superconducting string loop cusp
annihilates.  Because of the very large Lorentz factor achieved by the
contracting cusp when it is nearly at the point where it will trigger the
GRB, this beamed radiation is a very interesting feature of the model.  
It fits quite well with the current trend among GRB workers who claim that
recent observations provide strong evidence for some degree of beaming in
GRBs\cite{fenimore99}.

\begin{table*}[htb]
%\hline
%\hline
\caption{Inferred duration of the GRB (GW) emission phase
(giving a large part of the total energy) as a function of characteristic values
for the BATSE GRB fluence, and the assumed intergalactic magnetic field strength, with the CS parameter $\alpha \sim 0.4 \times 10^{-8}$. } 
\label{tbl-1}
\begin{tabular}{cccccc}
\multicolumn{1}{c}{\underline{$T^{GRBs}_{GWs}$ [s]} } &
\multicolumn{1}{c}{ } &
\multicolumn{1}{c}{\underline{Fluence S [erg cm$^{-2}$] } } &
\multicolumn{1}{c} {\underline{Magnetic Field B [G] }  } &
\multicolumn{1}{c} {\underline{Energy$^{Isot}_{\gamma}$ [erg] } }&
\multicolumn{1}{c} {\underline{GRB [BATSE]} } \\
\vspace{3pt} 
$6.7\times 10^{-6}$ & {   }   & $3\times 10^{-3}$ & $10^{-8}$\cite{battaner90} & $ \sim 3\times 10^{54}$ & { 971214 } \\ 
$3.2 \times 10^{-4}$ & { @ (Fig.1) }   & $1\times 10^{-4}$ & $10^{-8}$ & $ 6.7 \times 10^{53}$ & { 991216 } \\ 
$2.5 \times 10^{-3}$ & { * (Fig.1) }   & $1\times 10^{-5}$ & $10^{-7}$ & $ 6.7 \times 10^{53}$ & { "   " } \\ 
$6.7\times 10^{-3}$ & { XX (Fig.1)  }   & $3\times 10^{-4}$ & $10^{-7}$ & $\geq 3\times 10^{54}$ & { 990123 } \\ 
$2.0 \times 10^{-2}$ & {   }   & $1\times 10^{-5}$ & $10^{-7}$ & {    } & {   } \\
$200$ & {LISA Target}  & $1.5\times 10^{-8}$ & $10^{-7}$\cite{BHV2000} & $ \geq 5\times 10^{51}$ & {  } \\ 
\end{tabular} 
\end{table*}

%EDITOR: PLEASE PLACE TABLE 1 AROUND HERE.

Based on BATSE observations, we point out that the Berezinsky, Hnatyk and
Vilenkin SCCS scenario\cite{BHV2000} appears to be more well-motivated
than the earlier one for explaining GRBs from CSs. Thus, we shall follow
its main lines here in order to demonstrate that in such a view for the
central engine of GRBs, an accompanying burst of gravitational radiation
should also be released. As shown in Figs. 1 and 2, the characteristics of
such GW bursts make them potentially observable with the forthcoming
generation of Earth-based interferometric GW observatories LIGO, VIRGO, TAMA 
and GEO-600, and the space-borne LISA, as well as by the resonant-mass TIGAs.

A superconducting cosmic string loop, with energy per unit length $\mu
\sim \eta^2$ (where $\eta$ is the string symmetry-breaking scale),
oscillating in a magnetic field $B$, behaves like an $ac$ generator, and
an electric current $I_0 \sim e^2 B l$ flows in it.  Here $l \sim \alpha c
t$, defines the string loop invariant wavelength ($l \equiv E/\mu$, where
$E$ is the energy of the loop in the center-of-mass frame), and $\alpha
\sim \kappa_g G \mu << 1$ is a parameter determined by the gravitational
back reaction\cite{BHV2000}.

During brief time intervals, a noticeable augmentation of the local
current intensity can occur in domains quite close to the cusp location,
the point at which the string speed gets closest to the velocity of light.
Several cusps may appear during a single loop oscillation period.  Inside
a cusp domain $\delta l_c$ (at maximum contraction) the string shrinks by
a large factor, $l/\delta l$, leading to a relativistic Lorentz factor
$\Gamma \sim l/\delta l$ (in the string rest frame)\footnote{In the BHV2000 model this Lorentz factor may reach up to $\Gamma \sim 6.7 \times 10^7$. However, when the back-reaction effect of the electromagnetic radiation emitted in the process is taken into consideration the factor reduces to $\Gamma \leq 10^4$\cite{blanco-olum}, which is very consistent with the correspondings values inferred from BATSE observations\cite{meegan95,fenimore99}.}, this condition being
sustained for a timescale $\delta t_c \sim \delta l_c/c$, within a
physical length scale $\delta l_c \sim \delta l/\Gamma \sim l/\Gamma^2$.  
Most of the huge cusp rest energy is converted into kinetic energy.  
Since, in general, the string velocity is extremely high near to the cusp
collapse (it spreads out inside a cone with opening angle $\theta \sim
1/\Gamma$ oriented along the direction in which the cusp contracts), a
quadrupole distribution of the local energy density is expected to develop
(see Ref.\cite{BHV2000} for further details).

This scenario implies that a short burst of GW emission should occur in
the time leading up to cusp annihilation. In this brief time scale,
$\delta t_c$, the large asymmetric cusp shrinkage and energy reconversion
mean that a powerful GW burst would be emitted. As suggested above, we
expect that the GW burst and the $\gamma$-ray burst should have exactly
the same duration, and emphasize that long bursts ($\Delta t \sim 200$s)
may also be possible\cite{BHV2000}. This last point may be realised if a special combination of intergalactic magnetic strengths, i.e., $B \sim 10^{-7}$G, and GRB fluences, e.g., $S \sim 10^{-8}$erg cm$^{-2}$, is invoked. This possibility is illustrated in Table I, where several $\gamma$-ray fluences from particular events are combined with magnetic field strengths thought to exist around the GRB sources in the context of the SCCS cusp annihilation mechanism.

The GW characteristics (amplitude and frequency) can be estimated using
the typical dynamical timescale for GRBs in this model.  According to
BHV2000\cite{BHV2000}, the $\gamma$-ray timescale (GRB duration) is given
by

\begin{eqnarray}
T^{GRBs}_{GWs} & \simeq & 0.24 {\rm  ms}\; \left(\frac{B}{10^{-8} {\rm G}\cite{battaner90}}\right)^2 \left(\frac{\alpha}{10^{-8}} \right)^4 \nonumber \\ 
&  \times & \left(\frac{10^{-4} \; {\rm erg cm^{-2}} }{S }\right) 
 \frac{(1 + z)^{-1}}{[(1 + z)^{-1/2} - 1]^{-2}},\label{time} 
\end{eqnarray}

where $S$ is the GRB fluence in units of $10^{-4}$ erg cm$^{-2}$, $B$ is
the intergalactic magnetic field in units of $10^{-8}$ G\cite{battaner90}, and $z \sim 4$ is the source redshift. This value for the redshift $z$ at which the SCCS is located in this model has been taken in agreement with BATSE observations of the most distant and most energetic GRBs ever detected: GRB000131 at $ z = 4.5$, Andersen et al. (VLT Team)\cite{andersen}. It is also consistent with existing models for ultrahigh-energy cosmic rays and the observed shape of their spectrum.

This is a timescale that could be observed by BATSE (time resolution
$\Delta t \sim 100 \mu$s) in very short GRBs from this sort of
cosmological source (see Table I), were it not for its low efficiency for 
such bursts (see for example Trigger Number 01453, $\Delta t = 0.006 \pm 0.0002$ 
in Table 1 in Cline, Matthey and Otwinowski\cite{cline00}). In what follows we concentrate on this kind of GRB.

One can use the general relativity (GR) quadrupole formula to make an estimate of the characteristic GW amplitude: the dimensionless space-time strain ($h$),
generated by the non-spherical dynamics of the {\it cusp} kinetic energy,
is related to the distance of the CS ($D$) by $h_{ij} = \frac{2 G}{c^4 D}
\frac{d^2Q_{ij}}{dt^2}$, where $Q_{ij}$ is the second moment of the mass
distribution (quadrupole mass-tensor) in the transverse traceless (TT)
gauge and is given by $Q_{ij} \equiv \int \rho (x_i x_j - 1/3 x^2) d^3x$,
with $\rho$ being the mass-energy density of the source. This expression
can be rewritten as \cite{schutz99}

\be
h = \frac{2 G}{c^4} \frac{E_{Non-Sym}}{D_L}, \label{haga} 
\ee

where $D_L$ is the luminosity distance defined by $D_L = r_z (1 + z)$,
with $z$ being the source redshift (inferred from the spectrum of the GRB host), and $r_z$ is the comoving distance given by $r_z = 2
\frac{c}{H_0} {(1 - [1 + z]^{-1/2})}$. Here $H_0$ is the present-day value
of the Hubble constant. For the case which we are studying, we can
approximate the non-symmetric part of the kinetic energy of the
annihilating cusp as ${E_{Non-Sym}} \sim M_{c}
\left(\frac{dl}{dt}\right)^2 \label{energy}$, where $M_{c} \sim \mu \delta
l_c \sim \mu c \delta t_c$ is the total mass of the cusp.  At the time
when the gamma-ray outburst occurs, we can express the time derivative of
the cusp characteristic linear dimension as $\left(\frac{\delta l}{\delta
t}\right)^2 \sim \left(\frac{\Gamma \delta l_c}{\delta t_c}\right)^2 \sim
\Gamma^2 c^2$, with $\delta t \sim \delta t_c$, as discussed by BHV2000.  
We also identify the GW pulse duration as $\delta t_c \sim
T^{GRBs}_{GWs}$, given by Eq.(\ref{time}). Using these expressions,
we can write ${E_{Non-Sym}} \sim \mu c^3 \Gamma^2 \delta t_c$, and
Eq.(\ref{haga})  becomes

\begin{eqnarray}
h \sim \left[\frac{G}{c D_L}\right] \mu \Gamma^2  T^{GRBs}_{GWB}  & \sim & 1.9 \times 10^{-22}\Gamma^2  \left({T^{GRBs}_{GWB}}\right) \times \\ &  & \left(\frac{10^{28} {\rm \; cm}}{D_L}\right) \left(\frac{\mu}{(10^{16} \;{\rm GeV})^2}\right),  \nonumber
\end{eqnarray}

where we have used as the GUT symmetry breaking scale for the SCCS:
$\eta \sim 10^{16}$GeV, and have assumed that the source is at a distance equal to the Hubble radius. Consistently with current observations, and following
BHV2000, we will consider sources at low $z \sim 5$\cite{andersen} for
which $\Gamma = 300$ and $\Gamma = 10^3$ are plausible Lorentz factors
according to BATSE observations and the fireball model\cite{meegan95}. 
Recall that
high $\Gamma$ values are needed in order for the fireball to avoid
overproduction of electron-positron pairs\cite{piran98}. From Eq.(3) we
then have $h^{\Gamma=300}_{6.7 {\rm ms}} \sim 9.4 \times 10^{-21} {\rm
Hz}^{-1/2}$, $h^{\Gamma=300}_{2.4 {\rm ms}} \sim 2.1 \times 10^{-21}
{\rm Hz}^{-1/2}$ and $h^{\Gamma=10^3}_{0.32 {\rm ms}} \sim 1.5 \times
10^{-21} {\rm Hz}^{-1/2}$, for the burst duration and Lorentz factor, as indicated.

The maximum frequency of the GW burst in the reference frame of the loop
can be approximated as $f_{GW} \sim t_{dyn}^{-1}$, with $t_{dyn} \sim
T^{GRBs}_{GWs}$ being the dynamical timescale for annihilation of the
cusp. This then implies $h \sim f_{GW}^{-3/2}$ and so we can write

\be
f_{GW} \equiv {(T^{GRBs}_{GWs})^{-1}}  \sim  
\left(\frac{\delta l_c}{c}\right)^{-1} \sim 150, 420, 3200 \;{\rm Hz},
\ee

for GRBs with the durations given in Table I. Such frequencies fall just
within the range of highest sensitivity for the LIGO (I,II), VIRGO and
GEO-600 interferometers, and for the Brazilian M\'ario Sch\"onberg and Dutch Mini-GRAIL TIGAs, as shown in Figure 1. Thus, in the high frequency regime, the
bursts may be detectable for higher $\Gamma$ and much lower $\eta$, for a
given $\Delta t$, than is the case at lower frequencies.

\begin{figure}
%\vspace{9pt}
%\framebox{\rule[-21mm]{0mm}{43mm}
\includegraphics*[width=21pc]{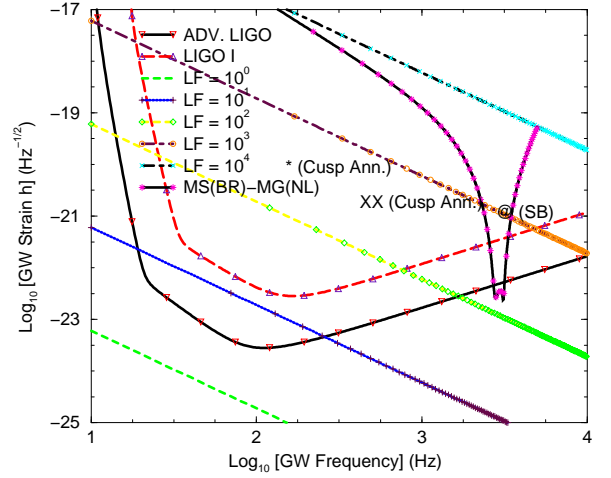} 
\caption{Locus of the GW characteristics (-3/2 slope equally spaced lines)
of the burst produced by the SCCS cusp annihilation (computed for $\eta = 10^{16}$GeV and $\Delta t =  2.4$ms) plotted against strain (burst) spectral densities and frequency bandwidth of the interferometers
LIGO(I,II) and the future {\it twin} TIGAs: the {\it M\'ario Sch\"onberg} (MS, Brazil) and the {\it Mini-GRAIL} (MG, Netherlands). Symbols *, XX and @ denote GW bursts with frequencies 150 Hz ($\Gamma$ = 300), 420 Hz ($\Gamma$ = 300) and 3200 Hz ($\Gamma = 10^3$, very short bursts (SB)), respectively. These GW signals may be triggered simultaneously with GRBs having the characteristics shown in Table I and Lorentz factors (LF) $\Gamma$ as indicated here.}
\end{figure}

%EDITOR: PLEASE PLACE FIGURE 1 AROUND HERE.

For the very low GW frequency band $10^{-4}-10^{0}$ Hz, the LISA antenna could observe these signals even for extremely low GUT energy scales but large $\Gamma$ factors, as shown in Figure 2 (of course, it could also observe the GW pulses for higher $\eta$ and lower $\Gamma$).

%EDITOR: PLEASE PLACE FIGURE 2 AROUND HERE.

\begin{figure}
%\vspace{9pt}
%\framebox[55mm]{\rule[-21mm]{0mm}{43mm}
\includegraphics*[width=21pc]{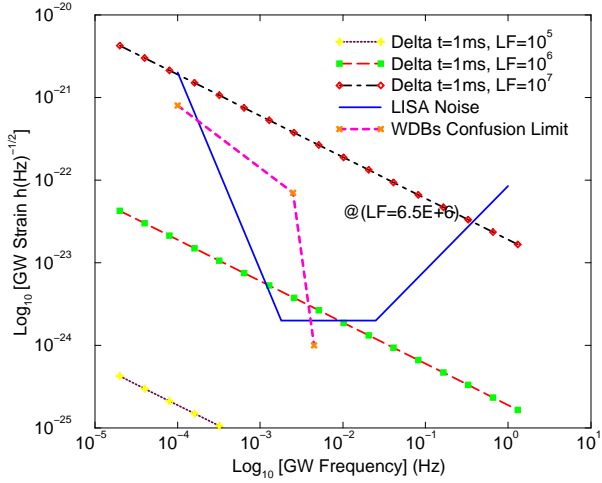} 
\caption{Locus of the GW characteristics (computed for $\eta = 10^{10}$ GeV and  $\Delta t = 1$ms) compared with the strain (burst) spectral density and the frequency bandwidth of LISA. The confusion limit produced by the  background of  white dwarf binaries is also plotted. The symbol @ indicates a GW burst with frequency $10^{-2}$ Hz and $h = 4\times 10^{-23}$, in units of (Hz)$^{-1/2}$, for a GRB with $\Gamma = 6.5 \times 10^6$.}
\end{figure}

%\subsection{Ultra High Energy Neutrino Production in the BHV2000 SCCS model of %GRBs}

In the context of the BHV2000 SCCS model for GRBs it is also possible for
an outburst of ultra high energy cosmic rays to be released simultaneously
with the GRB surge.  Assuming that a packet of such particles comes in the
form of a neutrino burst, we can estimate the neutrino time-of-flight
delay with respect to both the GRB and GW signals. This measurement can
provide an indication of the neutrino mass eigenstates. Next we make a rough estimate of the
overall characteristics of such a $\nu$-spectrum, and use it to constrain
the predicted time lag. (A more consistent calculation of the actual
UHECR spectrum in this picture, in the light of their detectability by the
AUGER experiment, is now under way\cite{nos}.)

Topological defects or unstable relic particles produce ultra high energy
photons at a rate $\dot{n}_p = \dot{n}_{p,0}(t/t_0)^{-m}$, where $m = 0,
3, 4$ for decaying particles, ordinary CSs and necklaces, and SCCS,
respectively\cite{vilenkin81,kibble94}. In the GRB fireball
picture\cite{piran98}, the detected $\gamma$-rays are produced via
synchroton radiation coming from ultrarelativistic electrons boosted by
internal shocks in an expanding relativistic blast wave (wind) consisting
of electron-positron pairs, some baryons and a huge number of photons. The
typical synchroton frequency is constrained by the characteristic energy
of the accelerated electrons and also by the intensity of the magnetic
field in the emitting region. Since the electron synchroton cooling time
is short compared with the wind expansion time, electrons lose their
energy radiatively. The standard energy of the observed synchroton photons
(see Refs.\cite{waxman,meszaros} for a more complete review of this
mechanism) is $E^b_\gamma = \frac{\Gamma \hbar \gamma^2_e e B}{m_e c}$
which is then given by

\begin{eqnarray}
E^b_\gamma & \simeq & 45 {\rm MeV}\; \xi^{1/2}_B \xi^{3/2}_e
\left(\frac{L_{\gamma}}{10^{54}\; {\rm erg s^{-1}} }\right)^{1/2}
\left(\frac{10^3}{\Gamma}\right)^{2} \nonumber \\ & \times &
\left(\frac{0.25\; {\rm ms}}{\Delta t}\right),
\end{eqnarray}

where $L_\gamma = 10^{54}$ ergs$^{-1}$ is the power released in the most energetic GRBs observed by BATSE. This $L_\gamma$ may imply Lorentz
expansion factors $\Gamma \sim 10^3$, and we assume this in the following.
$\Delta t = 0.25 {\rm ms}$ is the {\it inferred} timescale for the shorter
$\gamma$-ray bursts from the BHV2000 SCCSs, $\xi_B$ is the fraction of
the energy carried by the magnetic field: $4 \pi r^2_d c \Gamma^2 B^2 =
8\pi \xi_B L$, where $L$ is the total wind luminosity, and $\xi_e$ is the
luminosity fraction carried away by electrons. No theory is available to provide
specific values for $\xi_B$ and $\xi_e$. However, for values near to
equipartition, the break energy $E^b_\nu$ for photons in this model is in
agreement with the observed one for $\Gamma \sim 10^3$ and $\Delta t =
0.25$ ms, as discussed below. More precisely, the hardness of the GRB spectra, which extend up to 18 GeV, constrains the
wind to have Lorentz factor $\Gamma \sim 10^3$, while the observed
variability of the GRB flux on a timescale $\Delta t \leq$ 1ms implies
that the internal collisions occur at a distance from the center of $ r_d
\sim \Gamma^2 \Delta t$, due to variability of the central engine on the
same timescale. Since most of the BATSE observed GRBs show variability
with $\Delta t \leq 10$ ms and there is rapid variability with $\Delta t \leq 1$
ms, the implied characteristic size of the emitting region is $r_{em}
\sim 10^7$ cm which means that it must be a compact domain. The SCCS loop
cusp clearly satisfies this constraint.

In the acceleration region, protons (the fireball baryon load) are also
expected to be shocked. Their {\it photo-meson} interaction with observed
$\gamma$-burst photons should produce a surge of neutrinos almost
simultaneously with the GRB via the decay $p + ^{n_0}_\gamma
\longrightarrow \pi^+ \leftrightarrow \mu^+ + \nu_\mu \leftrightarrow e^+
+ \nu_e + \bar{\nu}_\mu + \nu_\mu$. The neutrino spectrum for a fireball
driven explosion is expected to follow the observed $\gamma$-ray spectrum, which is approximately a broken power-law $\frac{dN_\gamma}{dE_\gamma} \propto
E^{-\beta}_\gamma$, with $\beta \sim 1$ for low energies and $\beta \sim
2$ for high energies as compared with the observed {\it break energy} $
E^\beta_\gamma \sim 45$ MeV, where $\beta$ changes. The interaction of
protons from the surrounding medium, accelerated to a power-law
distribution
$\frac{dN_p}{dE_p} \propto E^{-2}_p$, with the fireball photons, leads to a
broken power-law neutrino spectrum $\frac{dN_\nu}{dE_\nu} \propto
E^{-\beta}_\nu$, with $\beta =1$ for $E_\nu < E^b_\nu$, and $\beta = 2$
for $E_\nu > E^b_\nu$. Thus the neutrino break energy $E^b_\nu$ is fixed
by the threshold energy of photons from {\it photo-production} interacting
with the dominant $\sim 45$ MeV fireball photons (in our case), and is

\begin{eqnarray}
E^b_\nu \simeq 1.3 \times 10^{15} \left(\frac{\Gamma}{10^3}\right)^{2}
\left(\frac{45\; {\rm MeV}}{E^b_\gamma}\right) {\rm eV}.\label{spectrum}
\end{eqnarray}

Thus, for $\nu$s produced with the above energy a further Fermi cycle in the ultrarelativistic blast wave
may amplify the UHECR energy by a factor of $\Gamma^2$ which, for the case
of protons, may push them over the GZK limit\cite{greisen} ($\nu$s do not
have a GZK cut-off). The part of the total fireball luminosity that escapes 
as  the $\nu$-flux is determined by
the efficiency of pion production. The energy fraction lost via pion
production by protons {\it producing $\nu$s above the break energy} is
essentially independent of the energy and can be expressed as

\begin{eqnarray}
f_\pi & = & 0.23 \left(\frac{L_\gamma}{10^{54} \;{\rm erg s^{-1}} }\right)
\left(\frac{45 \; {\rm MeV}}{E^b_\gamma}\right)
\left(\frac{10^3}{\Gamma}\right)^{4} \nonumber \\ & \times&
\left(\frac{0.25\; {\rm ms}}{\Delta t}\right).
\end{eqnarray}

Thus, an important part of the total wind energy is given to these very
high energy $\nu$s.

The time-of-flight delay of the $\nu$s with respect to the GWs (and the
GRB) may be computed by using the spectrum (Eq.\ref{spectrum}) and Table
II, above. This gives\cite{raffelt99}

\begin{eqnarray} 
\Delta T^{GWs-GRBs}_\nu & \sim & 1.545 \; {\rm s} {\label{delay}} \\ & \times &
\left(\frac{D}{3\; {\rm Gpc}}\right) \left(\frac{m{^2_\nu}}{100\; {\rm
eV}^2}\right) \left(\frac{100\;{\rm GeV}^2} { {\rm
E}^2_\nu}\right).\nonumber
\end{eqnarray}

This equation was originally derived as a way of estimating the {\it
time-of-flight lag} between massive neutrinos and massless ones, which
should travel at the speed of light\cite{raffelt99}. However, it
can also be applied to the problem which we are studying here, 
since we assume that the
GWs propagate at the velocity of light, as in GR.  It turns out that the
detection of such a neutrino pulse with a delay of approximately 1.5
seconds (for $E_\nu \sim 10^{10}$eV) after the GRB and GW outbursts
from the same source on the sky would make it possible to impose
tighter bounds on the neutrino mass spectrum, since the source distance
may be estimated from its redshift and the GWs detected. Of course,
there are some uncertainties involved in the derivation of Eq.(\ref{delay}). However, it is foreseeable that if atomic clocks were
installed in both the GW and $\nu$ observatories, a very precise
measurement of the arrival times might be obtained, making this
determination a plausible one in the near future.

To summarize, since the $\nu$-spectrum ranges over fourteen orders of magnitude (MeV $\nu$s from SN1987A, above GZK $\nu$s detected by AGASA, Fly's Eye, etc., see Table II), and the $\nu$-energy can be measured directly at the detector,
the detection of any species of neutrino in near spatial and temporal coincidence with observed GW + GRB
signals might yield a very accurate estimate of the time-delay
between them. Through the analysis of such a time lag, one may verify or
rule out the BHV2000 SCCS model, and also clarify the value of the GW
propagation velocity which is a quantity of great interest for
descriminating between different relativistic theories of
gravity\cite{herman,will}.

\begin{table}
%\hline
%\hline
\caption{Time delay between GWs (GRBs) and $\nu$-bursts as a
function of the $\nu$-energy in the BHV2000 SCCS model. } 
\label{tbl-2}
\begin{tabular}{cccccc}
\multicolumn{1}{c}{\underline{$\Delta T^{GRBs-GWs}_{\nu s}$ [s]} } &
\multicolumn{1}{c}{  } &
\multicolumn{1}{c}{  } &
\multicolumn{1}{c} {  } &
\multicolumn{1}{c} {  } &
\multicolumn{1}{c}{ \underline{$\nu$ Energy [eV]} } \\
\vspace{3pt} 
$1.545 $ & {   }   &   {   } & {   } & {   } & {   $10^{10}$  }  \\ 
$1.545 \times 10^{-8}$ & {   }   &   {   } & {   } & {   } & $10^{14}$   \\ 
$1.545 \times 10^{-20}$ & {   }   &  {   } & {   } & {   } & { $10^{20}$  }  \\ 
\end{tabular} 
\end{table}

%EDITOR: PLEASE PLACE TABLE 2 AROUND HERE.

{\bf Acknowledgements}: We are indebted to Prof. John C. Miller (SISSA and OXFORD) for his patience in reviewing our paper and for his important suggestions for making  this manuscript comprehensible. HJMC thanks the Centro Latino-Americano de F\'{\i}sica (Rio de Janeiro) and Conselho Nacional de Desenvolvimento e Pesquisa (CNPq, Brazil) for financial support. DMG also thanks CNPq for a Graduate Scholarship.


\begin{thebibliography}{99}

\bibitem{vilenkin81}A.  Vilenkin, Phys. Rev. Letts. 46, 1169 (1981). See
also A. Vilenkin and E. P. S. Shellard, {\it Cosmic strings and other
topological defects}, C.U.P., Cambridge, England (1994), and references
therein.

\bibitem{kibble94}M.  Hindmarsh and T.  W.  B.  Kibble, report
hep-ph/9411342 (1994).


\bibitem{MB92-93}R.  Brandenberger, A. T. Sornborger and M. Trodden, Phys.  
Rev.  D 48, 940 (1993). M. Mohazzab and R. Brandenberger, Int. J. Mod.
Phys. D2, 183 (1993).


\bibitem{BHV2000}V.  Berezinsky, B.  Hnatyk and A.  Vilenkin,
astro-ph/0001213 (2000).


\bibitem{meegan95}C.  Meegan,  et al., Astrophys.  J. Supp. 106, 65 (1995).



\bibitem{greisen}K. Greisen, Phys. Rev. Letts. 16, 748 (1966); Z. T.
Zatsepin and V. A. Kuz'min, P'sma Zh. Eksp. Teor. Fiz. 4, 144 (1966).




\bibitem{bhatta98}P.  Bhattacharjee, report hep-ph/9811011 (1998).

\bibitem{totani99}T.  Totani, Mon. Not. R. Astr. Soc. 307, L41 (1999).

\bibitem{allen00}B. Allen and A. C. Ottewill, report gr-qc/0009091 (2000). 

\bibitem{babul87}A.  Babul, B. Paczy\'nski and D.  Spergel, Astrophys. J.
Letts. 316, L49 (1987).

\bibitem{fenimore99}E. E. Fenimore and E. Ramirez-Ruiz, Astrophys. J. 518,
375 (1999).

\bibitem{blanco-olum}J. J. Blanco-Pillado and K. Olum, report astro-ph/0008297 (2000).

\bibitem{battaner90}E. Battaner, E. Florido and M. I. Sanchez-Saavedra,
Astron. Astrophys. 236, 1-8 (1990).

\bibitem{andersen}Andersen, M., et al., ``VLT Identification of the optical afterglow of the gamma-ray burst GRB000131 at $z = 4.5$", to appear in Astron. \& Astrophys. Letts., Dec. 1 (2000).

\bibitem{cline00}D. B. Cline, C. Matthey and S. Otwinowski, Astrophys. J. 527, 827 (2000).

\bibitem{schutz99}B. F. Schutz, in Proceedings of the Como Conference on
{\it Gravitational Waves in Astrophysics, Cosmology and String Theory},
April 18-23, Como, Italy (1999), to be published.


\bibitem{piran98}M. J. Rees and P. M\'esz\'aros, MNRAS, 258, L41 (1992).
T. Piran, Phys. Reports 314, 575 (1999).

\bibitem{nos}H. J. Mosquera Cuesta and D. Morej\'on Gonz\'alez, in preparation.


\bibitem{waxman}E. Waxman, report astro-ph/0002243 (2000) and Refs. therein.

\bibitem{meszaros}P. M\'esz\'aros \& J. N. Bahcall, report hep-ph/0004019
(2000).


\bibitem{raffelt99}G.  Raffelt, {\it Stars as laboratories for fundamental
physics: The astrophysics of}..., Univ. of Chicago Press (1996).

\bibitem{herman} H. J. Mosquera Cuesta, M. Novello \& V. A. De Lorenci,
submitted to Phys. Rev. D, May (2000).


\bibitem{will}C. M. Will,  Phys. Today 38, October (1999).




\end{thebibliography}
\end{document}